\documentstyle[aps,epsf,prl,multicol]{revtex}
\draft

\begin{document}
\title{Square patterns in Rayleigh-B\'enard convection with rotation about
a vertical axis} 
\author{Kapil M. S. Bajaj, Jun Liu, Brian Naberhuis, and Guenter Ahlers}
\address{Department of Physics and Center for Nonlinear Science, University
of California, Santa Barbara, California 93106, USA} \date{\today} \maketitle
\begin{abstract}
We present experimental results for Rayleigh-B\'{e}nard convection with
rotation about a vertical axis at dimensionless rotation rates $0 \le \Omega
\le 250$ and $\epsilon \equiv \Delta T / \Delta T_c - 1 \alt 0.2$. Critical
Rayleigh numbers and wavenumbers agree with predictions of linear stability
analysis. For $\Omega \agt 70$ and small $\epsilon$ the patterns are cellular
with local four-fold coordination and differ from the theoretically expected 
K\"{u}ppers-Lortz-unstable state. Stable as well as intermittent defect-free
square lattices exist over certain parameter ranges. Over other ranges defects
dynamically disrupt the lattice but cellular flow and local four-fold
coordination is maintained.

\end{abstract}

\begin{multicols}{2}

\begin{figure}[t]
\epsfxsize=2.75in \centerline{\epsffile{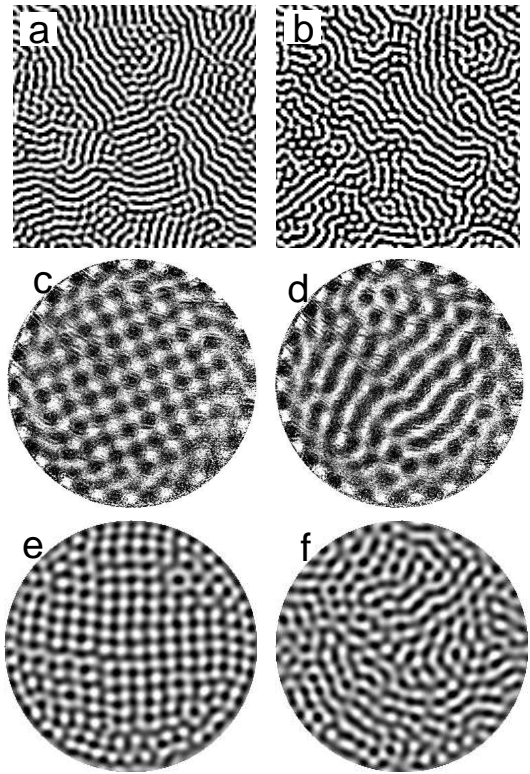}}
\vskip 0.1in
\caption{Shadowgraph images of convection patterns viewed from above for
CO$_2$ (a and b, only parts of patterns are shown; adapted from Ref.
\protect{\cite{HPAE98}}), water (c and d, the entire cell is shown), and Argon
(e and f, 90\% of the cell radius is shown) with ($\sigma$, $\Omega$,
$\Gamma$) equal to (0.93, 19.8, 40), (5.4, 170, 4.8), and (0.69, 181, 8.3)
respectively. 
For small $\Omega$ (a,b) the patterns have domains which are typical of the KL
instability for all $\epsilon$ near zero, as illustrated for $\epsilon = 0.06$
(a) and 0.18 (b). At larger $\Omega$, square patterns occur close to onset
[(c) $\epsilon=0.09$ and (e) $\epsilon=0.04$], but states similar to the KL
domains are observed for $\epsilon \protect\agt 0.1$ [$\epsilon=0.12$ in (d)
and $\epsilon=0.13$ in (f)]. In c and d, the cells along the periphery are the
wall mode.}
\label{fig:squares}

\end{figure}

The elucidation of spatio-temporal chaos (STC) remains one of the major 
tasks in the study of pattern formation in nonlinear 
dissipative systems.\cite{CH93} The best opportunities 
for theoretical understanding of experimental observations of a chaotic 
state exist when the mean-square amplitude of STC evolves 
continuously (i.e. via a supercritical 
bifurcation) from a spatially-uniform ground state; for in that case 
it should be possible at least in principle to derive from the equations of
motion 
of the system a systematic, simplified description in the form of 
Ginzburg-Landau equations. However, experimentally accessible supercritical 
bifurcations from the uniform state to STC are rare because most systems with
supercritical primary bifurcations  
become variational as the threshold is approached from above and thus 
approach a time-independent state. Convection in a shallow 
horizontal layer of a fluid heated from below (Rayleigh-B\'enard 
convection or RBC) and rotated 
about a vertical axis is one of the exceptions. Convection occurs when the
temperature difference $\Delta T$ exceeds a critical value $\Delta
T_c(\Omega)$ ($\Omega$ is the rotation frequency) and leads to a velocity
field ${\bf v}$. The Coriolis force 
$\Omega \times {\bf v}$  renders the system non-variational even close to 
the convective threshold and thus permits the existence of STC at onset. 
K\"uppers and Lortz\cite{KL69} (KL) predicted that a primary supercritical
bifurcation 
leads to a state of unstable convection rolls provided 
$\Omega > \Omega_{KL}$. In this KL state, rolls of one orientation are 
unstable with respect to another set of rolls with an angular orientation
relative 
to the first which is advanced in the direction of rotation by an angle
$\theta$.\cite{KL69,K70,CB79}
The new set, however, is equally unstable to yet another, and so forth.
Several experiments confirmed the existence of the KL state at relatively
large $\epsilon \equiv \Delta T/\Delta T_c -1$.\cite{He79,BH80,ZES91} Recent
experiments \cite{BCEHLA92,HEA95,HPAE98} have shown its existence close to
threshold for $\Omega_{KL} \simeq 12 \alt \Omega \alt 20$ and have verified
the supercritical nature of the bifurcation.  The instability was observed to
lead to a chaotically time dependent co-existence of domains with different
roll orientations. Two examples\cite{HPAE98} of this domain chaos are shown in
Figs. \ref{fig:squares}a and b. 

\narrowtext

Theoretically it is expected \cite{KL69,K70,CB79,Bu98} that the nature of the
bifurcation and of the nonlinear state above it should remain qualitatively
unchanged as $\Omega$ is increased above the range explored previously by
experiment. The measurements which we report here show that this is not the
case. For $\Omega \equiv 2\pi f d^2/\nu \agt 70$ ($f$ is the rotation
frequency in Hz and $\nu$ the kinematic viscosity) we find that the
bifurcation does remain supercritical as predicted, but that the convection
pattern above onset has no similarity to the expected KL state. Instead the
pattern consists of cells which are usually arranged so as to have local
four-fold coordination. Over significant parameter ranges the cells
``crystallize" and form a square lattice.\cite{Heikes} Typical examples are
shown in Figs.~\ref{fig:squares}c and e. Depending on parameter values, the
lattice can be stable, can be intermittently disrupted by defects which seem
to be injected from the boundary, or can be continuously disorderd with many
defects within it (maintaining, however, the cellular character with
predominantly local four-fold coordination). At larger $\epsilon$ ($\epsilon
\agt 0.1$), we do observe patterns reminiscent of the KL state, as illustrated
in Figs.~\ref{fig:squares}d and f. The existence of the square state for
$\epsilon$-values below those where the KL state is observed is contrary to
theoretical prediction. This disagreement with theory occurs in a parameter
range where the stability analysis was thought to be reliable, and thus
presents a significant unresolved problem in the field of pattern formation.
It also implies that the use of the KL state as a prototype for STC at onset
is valid only at relatively small $\Omega$.
    
We used water and Argon in two different apparatus \cite{ACBS94,DBMTHCA96}.
The Prandtl numbers $\sigma \equiv \nu/\kappa$ ($\kappa$ is the thermal
diffusivity) were 5.4 and 0.69 respectively. The temperatures of the tops of
the cells were regulated to $\pm 1$mK at
30.07 $^\circ$~C (water) and at 37.5 $^\circ$~C (Argon). 
Aspect ratios $\Gamma \equiv r/d = 4.8$ and 8.3 ($r$ is the radius and $d$ the
height of the fluid layer) were used for water 
($r=38.1$~mm,  $d=7.9$~mm) and Argon ($r = 33.25$~mm, $d= 4.0$~mm)
respectively.
The apparatus were rigidly mounted on rotating tables capable of
rotation up to $f = 1$~Hz, covering the range $0 \leq \Omega \leq 250$.
Shadowgraph assemblies were mounted ontop of the convection apparatus to
obtain images in the rotating frame. For Argon, the pressure was controlled to
$\pm 2$mbar. The experiments were carried out by raising the bottom-plate
temperatures at fixed $\Omega$.  Both heat-transport measurements and
shadowgraph images were taken after waiting for at least a horizontal
diffusion time $\tau_h=\Gamma^2 d^2/\kappa$.  Onset was determined both from
the heat transport and from amplitudes of the  shadowgraph images.

The dimensionless control parameters are the Rayleigh number $R \equiv \alpha
g d^3 \Delta T/\kappa \nu$ and $\Omega$. Here $\alpha$ is the isobaric thermal
expansion
coefficient, $g$ the acceleration due to gravity, and $\Delta T$ the
temperature difference across the cell. The effect of centrifugal force given
by the ratio of centrifugal-to-gravitational force $F = 4\pi^2 f^2 r/g$ 
is less than 0.03 in most of the experiments and maximally 0.12 for a few
runs. 
Since $\Omega$ is scaled by $\tau_\nu \equiv d^2/\nu$, increasing the pressure
$P$ 
of Argon increased $\Omega$ because it decreased $\nu$. We used this to vary
$\Omega$ by almost a factor of 4 at fixed $f$ (and thus $F$). 
Increasing $P$ changed $\sigma$ only from 0.68 at 20 bar to 0.695 at 40 bar.
It decreased the parameter $Q$ \cite{Bu67} which describes the extent of
departures from the Boussinesq  approximation. For Argon $Q$ ranged from 
0.03 at 20 bar to 0.008 at 40 bar for the highest $\Omega$ (largest $\Delta
T$). For water $Q \simeq -0.02$ at large $\Omega$. Since $Q$ was always small,
we expect non-Boussinesq effects to be negligible.   

\begin{figure}[t]
\epsfxsize=2.5in \centerline{\epsffile{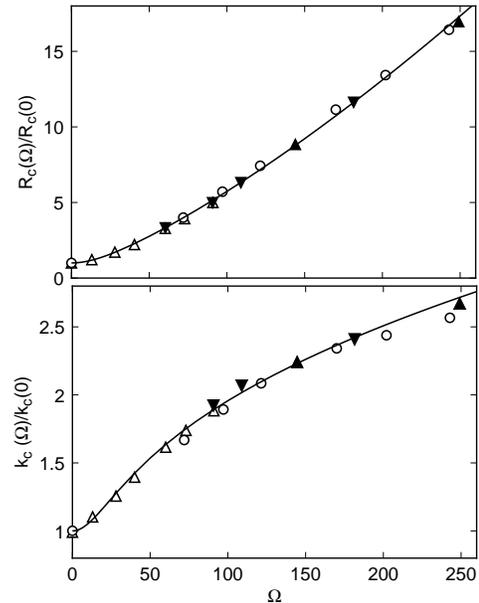}}
\vskip 0.2in
\caption{The critical Rayleigh numbers $R_c(\Omega)/R_c(0)$ (top) and
wavenumbers $k_c(\Omega)/ k_c(0)$ (bottom) as a function of $\Omega$  for
water ($\sigma=5.4$, open circles) and Argon at 20~bar ($\sigma=0.68$, open
triangles), 30 bar ($\sigma=0.69$ inverted solid triangles), and 40~bar
($\sigma=0.695$, solid triangles).  The results agree with the predictions for
a laterally infinite system (solid lines) \protect\cite{Ch53}. }
\label{fig:Randk}
\end{figure}

Figure~\ref{fig:Randk} summarizes measurements of the critical Rayleigh
numbers $R_c(\Omega)$ and wavenumbers $k_c(\Omega)$ for the onset of
convection in the sample interiors (the ``bulk mode"). The results agree 
with predictions based on linear stability analysis for a laterally infinite
system.~\cite{Ch53} These predictions are independent of $\sigma$ and are
given by the solid lines in Figs.~\ref{fig:Randk}. For $\Omega \agt 70$, the
bulk mode is preceded by a ``wall mode" consisting of a wave traveling (in the
rotating frame) in the direction opposite to $\Omega$.\cite{Wallmode} The wall
mode persists above $R_c$, and can be seen in Figs.~\ref{fig:squares}c and d.
The primary object of the present paper is the bulk mode. 
  
\begin{figure}[t]
\epsfxsize=2.75in \centerline{\epsffile{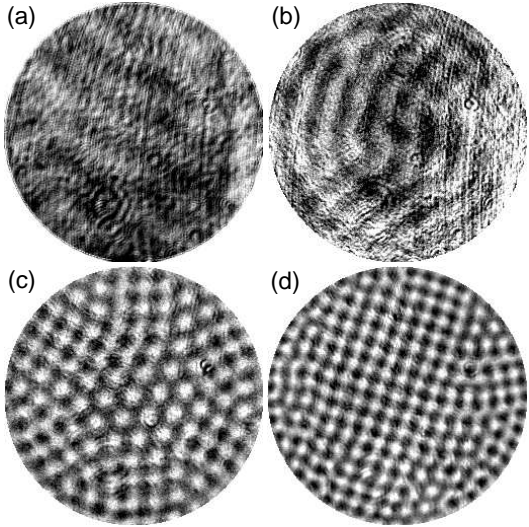}}
\vskip 0.1in
\caption{Patterns for Argon and various $\Omega$ at $\epsilon \simeq 0.04$.
$\Omega$ values are (a) 13, (b) 40 (c) 73 and (d) 145. Only 90\% of the cell
radius is shown. c and d are at the same physical rotation rate and at
different pressures.}
\label{AllW}
\end{figure}

Figure~\ref{AllW} shows shadowgraph images for $\epsilon=0.04$ for various
$\Omega$ in Argon. At small $\Omega$ (Fig.~\ref{AllW}a) the contrast of the
image is poor because $\Delta T$ and the wavenumber were small. Nonetheless
one can see that almost the entire cell was occupied by a single domain of
rolls. However, the time dependence of the pattern clearly revealed a dynamics
characteristic of the KL state.\cite{HEA95,HPAE98} 
At somewhat larger $\Omega$ (Fig.~\ref{AllW}b) several KL domains exist
simultaneously in the cell. These observations are consistent with previous
work\cite{HEA95,HPAE98} using larger $\Gamma$ and $\Omega \alt 20$. This
situation changes when $\Omega$ is increased beyond about 70, as illustrated
by Figs.~\ref{AllW}c and d. For $\Omega = 73$, 
the pattern is made up of cells rather than rolls. There are large domains
where the pattern has square symmetry. Six domains of squares meet in the cell
center and locally enforce a six-fold symmetry, 
maintaining however the cellular character of the flow. At larger $\Omega =
145$ the pattern consists of a square lattice, with some domain walls and
defects (right part of Fig.~\ref{AllW}d) apparently provoked by the cell wall.

\begin{figure}[t]
\epsfxsize=2.75in \centerline{\epsffile{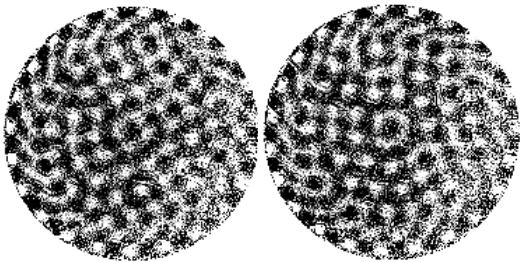}}
\vskip 0.1in
\caption{Disordered cellular patterns for $\sigma = 5.4$, $\Omega = 170$, and
$\epsilon = 0.052$. The entire cell is shown. Along its periphery is the wall
mode.}
\label{disorder}
\end{figure}

\begin{figure}[t]
\epsfxsize=2.75in \centerline{\epsffile{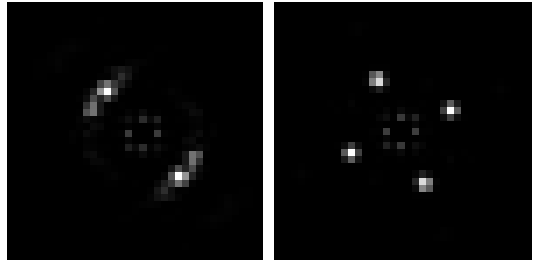}}
\vskip 0.1in
\caption{Structure functions of images 1c (right) and 1d (left).}
\label{S_of_k}
\end{figure}

There are ranges of $\Omega, \epsilon$, and $\sigma$ over which the patterns
near onset were dynamically disrupted by many defects, although cellular flow,
often with local four-fold coordination, was maintained. Two examples for
$\sigma = 5.4$, $\Omega = 170$, and $\epsilon = 0.052$ are shown in
Fig.~\ref{disorder}. In order to study the pattern dynamics in greater detail,
we examined the time evolution of the structure function $S({\bf k})$ (the
square of the modulus of the Fourier transform of the pattern). Two examples
of $S({\bf k})$ are shown in Fig.~\ref{S_of_k}.        

\begin{figure}[t]
\vskip -0.2in
\epsfxsize=2.75in \centerline{\epsffile{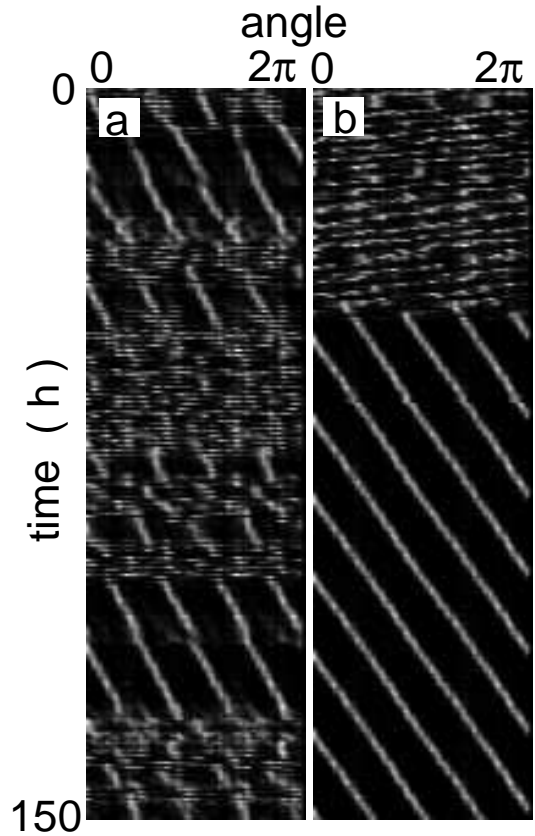}}
\vskip 0.1in
\caption{Angle-time plot for the radial average of the structure function for
$\sigma=5.4$ at $\Omega=170$ for (a) $\epsilon = 0.072$ and (b) $\epsilon =
0.091$. (a) is a representative section of a long run. The system
intermittently changes between a perfect square lattice and a disordered
pattern with local four-fold coordination. For (b) $\epsilon$ was stepped from
0.14 where the pattern was disordered (KL like) to 0.091 at time $t = -3$
hours. After about 16~$\tau_h$, the system spontaneously chose
the perfect-square pattern which then persisted for more than $38\tau_h$.}
\label{h2oang}
\end{figure}

\noindent The left one is typical of the KL state, whereas the right one
represents a square lattice. For $\Omega = 170$ and $\sigma = 5.4$, the radial
average $S(\Theta)$ of $S({\bf k})$ is shown in Fig.~\ref{h2oang} as a
function of time. Figure~\ref{h2oang}a is a section of a long run at $\epsilon
= 0.072$; we believe it is representative of a statictically stationary
process. Disordered regions alternate irregularly with well ordered square
lattices. We believe the disorder is provoked by defects which are injected by
interaction with the wall mode, but further work is required to examine the
mechanism. Figure~\ref{h2oang}b is the result of an experiment where the
system was kept in the KL regime at $\epsilon = 0.14$ for a long time, and
where at time $t = -3$ hours $\epsilon$ was reduced to 0.091. The system
remained KL-like for nearly 44 hours (about 16 horizontal thermal diffusion
times), and then {\it suddenly} crystalized into a square lattice which
remained stable thereafter.

\begin{figure}[t]
\epsfxsize=2.75in \centerline{\epsffile{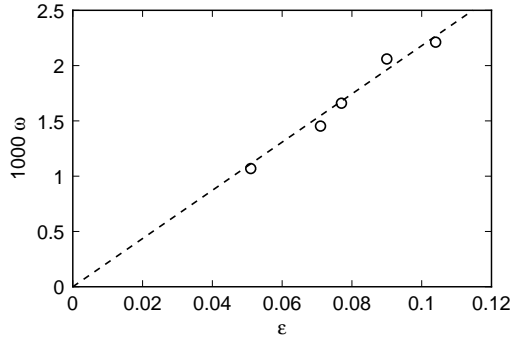}}
\vskip 0.1in
\caption{The rotation rate $\omega$ of the square lattice for $\Omega = 170$
and $\sigma = 5.4$. Time is scaled by $d^2/\nu$.}
\label{omega}
\end{figure}

It is interesting to note that the square lattice, whenever it exists, rotates
at a rate $\omega$ relative to the convection cell in the direction of the
overall rotation $\Omega$, albeit extremely slowly. Results for $\omega$ are
given in Fig.~\ref{omega}. The measurements suggest that $\omega$ vanishes as
$\epsilon$ goes to zero, as indicated by the dashed line. There is of course
no theoretical prediction for $\omega$, since even the existence of the
lattice is not understood.
For comparison, under similar conditions the wall mode is travelling in the
opposite direction at the much faster angular velocity of about $-0.15
\nu/d^2$, nearly independent of $\epsilon$. We believe that the interaction
between the fast-moving wall mode and the slow-moving bulk mode leads to the
defect formation which tends to disrupt the square lattice for some parameter
ranges.

Stable square patterns occur under other conditions in RBC. Convection of a
pure fluid in a cell with insulating top and bottom plates \cite{insulating}
and binary-fluid convection with small Lewis numbers and a positive separation
ratio \cite{mixture_squares} are two such systems. However, in those cases the
critical wavenumber is reduced. Our critical wavenumbers agree with the linear
stability analysis for rotating RBC of a pure fluid with perfectly conducting
top and bottom boundaries. Our top and bottom boundaries have conductivities
which are orders of magnitude larger than those of the fluids. Our Argon
samples, even if they were contaminated with another component, would have
Lewis numbers of order one \cite{LA97} and thus would not produce a square
pattern due to mixture effects. Clearly impurities or poorly conducting
boundaries cannot be invoked to explain our observations.
Thus the occurrence of square patterns in rotating RBC at small $\epsilon$ is
unexplained, and we conclude that a
real discrepancy exists between theory\cite{KL69,K70,CB79} and experiment. 

We are grateful for stimulating discussions with F.H. Busse and W. Pesch.
This work was supported by the U.S. Department of Energy Grant
DE-FG03-87ER13738.

\end{multicols}

\begin{references}
\bibitem{CH93} M.~C. Cross and P.~C. Hohenberg, Rev. Mod. Phys. {\bf 65}, 851
(1993).

\bibitem{KL69} G.  K\"{u}ppers and D. Lortz, J. Fluid Mech. {\bf 35},
	609 (1969).

\bibitem{K70} G. K\"{u}ppers, Phys. Lett. {\bf 32A} 7(1970).

\bibitem{CB79}R. Clever and F.H. Busse, J. Fluid Mech. {\bf 94}, 609 (1979).

\bibitem{He79} H. E. Heikes, Ph. D. Dissertation, University of California, 	Los Angeles, unpublished, 1979.

\bibitem{BH80}F. H. Busse and K. E. Heikes, Science {\bf 208}, 173 (1980)

\bibitem{ZES91} F. Zhong, R. E. Ecke and V. Steinberg, Physica D 
	{\bf 51}, 596 (1991).

\bibitem{BCEHLA92} E. Bodenschatz, D.S. Cannell, R.Ecke, Y.C. Hu, K. Lerman,
and G. Ahlers, {\it Physica D} {\bf 61}, 77 (1992).

\bibitem{HEA95}Y.-C. Hu, R.E. Ecke, and G. Ahlers, Phys. Rev. Lett. {\bf 74},
5040 (1995).

\bibitem{HPAE98}Y.-C. Hu, W. Pesch, G. Ahlers, and R.E. Ecke, Phys. Rev. E,
unpublished.

\bibitem{Bu98} F. Busse, private communication.

\bibitem{Heikes} A square pattern near onset was observed indirectly by Heikes
(Ref. \cite{He79}). However, that work did not have the optical resolution
needed for direct observations. Instead, the system was equilibrated at small
$\epsilon$, and then stepped to $\epsilon = {\cal O}(1)$. The rapidly growing
amplitude then became observable, and suggested that the pattern at the
previous steady state had been one of squares. To our knowledge these
observations were never reported in the published literature.

\bibitem{ACBS94}G. Ahlers, D.S. Cannell, L.I. Berge, and S. Sakurai, {\it
Phys. Rev. E} {\bf 49}, 545 (1994).

\bibitem{DBMTHCA96} J.R. deBruyn, E. Bodenschatz, S. Morris, S. Trainoff,
Y.-C. Hu, D.S. Cannell, and G. Ahlers, {\it Rev. Sci. Instrum.} {\bf 67}, 2043
(1996).

\bibitem{Bu67}F. Busse, J. Fluid Mech. {\bf 30}, 625 (1967).

\bibitem{Ch53}S. Chandrasekhar, Proc. R. Soc. London {\bf A217}, 306 (1953);
{\em Hydrodynamics and Hydromagnetic Stability}, Oxford University Press,
London, 1961.

\bibitem{Wallmode}F. Zhong, R.E. Ecke, and V. Steinberg, Phys. Rev. Lett. {\bf
67}, 2473 (1991); J. Fluid Mech. {\bf 249}, 135 (1993); R. E. Ecke, F. Zhong
and E. Knobloch, Europhys. Lett. {\bf 19}, 177 (1992); L. Ning and R.E. Ecke,
Phys. Rev. E {\bf 47}, 3326 (1993); Y. Liu and R.E. Ecke, Phys. Rev. Lett.
{\bf 78}, 4391 (1997).

\bibitem{insulating}F.H. Busse and N. Riahi, J. Fluid Mech. {\bf 96}, 243
(1980); M.R.E. Proctor, J. Fluid Mech. {\bf 113}, 469 (1981).

\bibitem{mixture_squares}E. Moses and V. Steinberg, Phys. Rev. A {\bf 43}, 707
(1991); M.A. Dominguez-Lerma, G. Ahlers, and D.S. Cannell, Phys. Rev. E {\bf
52}, 6159 (1995).

\bibitem{LA97} J. Liu and G. Ahlers, Phys. Rev. E {\bf 55}, 6950 (1997).
\end{references}
\end{document}